\documentclass[letterpaper, 11pt]{article}
\pdfoutput=1

\usepackage[nocompress]{cite}
\usepackage{jheppub}

\usepackage{graphicx}
\usepackage{epstopdf}
\usepackage{amsmath, amssymb}

\usepackage{bm}
\usepackage{caption}
\usepackage{subcaption}

\usepackage{multirow} %package for putting different rows in different columns of a table
\usepackage{longtable} %package to allow multipage tables

\usepackage{romannum}%For roman numerals in the title

\usepackage[us,12hr]{datetime} 

\newcommand{\be}{\begin{eqnarray}}
\newcommand{\ee}{\end{eqnarray}}

\newcommand{\bn}{\begin{enumerate}}
\newcommand{\en}{\end{enumerate}}

\def\CA{{\cal A}}
\def\CB{{\cal B}}
\def\CC{{\cal C}}
\def\CD{{\cal D}}
\def\CE{{\cal E}}

\def\CI{{\cal I}}
\def\CJ{{\cal J}}

\def\CN{{\cal N}}

\def\CS{{\cal S}}
\def\CT{{\cal T}}

\def\G{\Gamma}

\def\half{\frac{1}{2}}

\def\Tr{{\rm Tr}}
\def\tr{{\rm Tr}}

\def\PE{\textrm{PE}}

\def\CA{{\cal A}} 
\def\CB{{\cal B}}
\def\CC{{\cal C}}
\def\CD{{\cal D}}
\def\CE{{\cal E}}

\def\CI{{\cal I}}
\def\CJ{{\cal J}}

\def\CN{{\cal N}}

\def\CS{{\cal S}}
\def\CT{{\cal T}}

\def\G{\Gamma}

\def\half{\frac{1}{2}}

\def\ft{\mathfrak{t}}

\def\Tr{{\rm Tr}}
\def\tr{{\rm Tr}}

\def\PE{{\rm PE}}

\title{Vanishing short multiplets in rank one 4d/5d SCFTs}

\author{Jaewon Song}
\affiliation{Department of Physics, Korea Advanced Institute of Science and Technology\\
291 Daehak-ro, Yuseong-gu, Daejeon 34141, Korea}
\emailAdd{jaewon.song@kaist.ac.kr}

\abstract
{
We study the short multiplet spectrum in 4d $\mathcal{N}=2$ superconformal theories of low rank using the full superconformal indices and the selection rules from the superconformal representation theory. We find a universal expression for the leading terms for the superconformal index of rank one $H_0, H_1, H_2, D_4, E_6, E_7$ theories. From this result, we argue that certain short multiplets appear in the operator product expansions involving stress-tensor, conserved current, and Coulomb branch operator vanish. We also apply the same procedure to 5d superconformal theories and find that $E_1$ theory has vanishing short multiplets analogous to that of the $H_1$ theory. 
}

\begin{document}
\maketitle

\section{Introduction and Summary}
Superconformal field theories (SCFT) have been a fruitful arena to discover new aspects of strongly-coupled quantum field theories. Moreover, a high degree of symmetry often allows us to develop novel tools to improve our quantitative understanding. In general, SCFTs may not have known Lagrangian descriptions, as in the case of the class $\CS$ theories \cite{Gaiotto:2009we,Gaiotto:2009hg} or Argyres-Douglas type theories \cite{Argyres:1995jj, Argyres:1995xn, Xie:2012hs}.

In this paper, we focus on four-dimensional $\CN=2$ superconformal theories of rank 1, labeled by Deligne-Cvitanovich exceptional series $H_0 \subset H_1 \subset H_2 \subset D_4 \subset E_6 \subset E_7 \subset E_8$. Here the rank is defined as the dimension of the Coulomb branch the theory has. These theories are also referred to as F-theory SCFTs since they are easily realized by a single D3-brane probing F-theory singularities labeled by the Kodaira classification \cite{Minahan:1996fg, Minahan:1996cj}.\footnote{All possible rank 1 theories are classified via its Seiberg-Witten geometry \cite{Argyres:2015ffa, Argyres:2015gha,Argyres:2016xua,Argyres:2016xmc, Argyres:2016yzz, Argyres:2017tmj}. See also \cite{Apruzzi:2020pmv} for another F-theoretic realization.} Following the widely believed conjecture that there is no rank zero $\CN=2$ SCFT, they qualify as the simplest theories among the possible $\CN=2$ theories in $4d$. Most notably, the original $H_0$ Argyres-Douglas theory saturates the lower bound on the central charge $c$ \cite{Liendo:2015ofa}. Under the SCFT/VOA correspondence \cite{Beem:2013sza}, the $H_0$ theory is mapped to the Yang-Lee minimal model, which is the simplest non-unitary chiral algebra. It has been further noticed that Argyres-Douglas (AD) theory and its generalizations have rather simple corresponding chiral algebras \cite{Buican:2015ina, Cordova:2015nma, Buican:2015tda, Song:2015wta, Lemos:2015orc, Cecotti:2015lab, Buican:2016arp, Xie:2016evu, Song:2016yfd, Fredrickson:2017yka, Cordova:2017mhb, Buican:2017uka, Song:2017oew, Beem:2017ooy, Fluder:2017oxm, Choi:2017nur, Buican:2017rya, Kozcaz:2018usv, Creutzig:2018lbc, Buican:2019huq, Watanabe:2019ssf, Foda:2019guo}, which suggests that they can be regarded as a natural analog of minimal models in 4d. Therefore they serve as a natural target for the superconformal bootstrap program \cite{Beem:2014zpa, Lemos:2015awa, Cornagliotto:2017snu, Gimenez-Grau:2020jrx}. 

In the paper \cite{Song:2017oew, Agarwal:2018zqi}, it was demonstrated that by combining the information of the Schur or Macdonald index (both of which can be obtained via associated VOA) and the selection rule for the OPE that can be obtained via superconformal characters (which was possible thanks to the work of \cite{Dolan:2002zh, Cordova:2016emh}), one could argue that certain OPE coefficients vanish for various generalized AD theories (see also \cite{Kiyoshige:2018wol}). It was also known that the rank 1 SCFTs, including the Minahan-Nemeschansky $E_n$ theories, also have the property of vanishing of OPE coefficients \cite{Beem:2013sza}. This analysis was possible since Macdonald indices for these theories were available, without which there would be ambiguity on interpreting each term in the index. However, the Schur sector does not capture operators in the Coulomb branch or any possible mixed OPE between the Higgs and the Coulomb branch operators. In the current paper, we would like to expand the previous analysis beyond the Schur sector so that we have access to the other universal part in the $\CN=2$ SCFT, namely the Coulomb branch. 

Apart from the $D_4$ theory, all the rank 1 SCFTs (including the ones that are not in the exceptional series) do not have a Lagrangian description with manifest $\CN=2$ supersymmetry, which posed a challenge to compute the superconformal index beyond the Schur sector. However, it has been found that there exist $\CN=1$ Lagrangian theories that flows to some of the rank 1 SCFTs, namely $H_0, H_1, H_2, E_6, E_7$ \cite{Gadde:2010te, Gadde:2015xta, Maruyoshi:2016tqk, Maruyoshi:2016aim, Agarwal:2016pjo, Agarwal:2018ejn}.\footnote{$\CN=1$ Lagrangian description for other non-Lagrangian $\CN \ge 2$ theories are given in \cite{Agarwal:2017roi, Benvenuti:2017bpg, Razamat:2019vfd, Zafrir:2019hps, Razamat:2020pra, Zafrir:2020epd}.} This allows us to compute the full superconformal indices for these theories. We would like to combine the explicit expression for the index with the OPE selection rules involving universal short multiplets in $\CN=2$ theory, namely stress-tensor ($\CT$), conserved current ($\CJ$), and the Coulomb branch (or $\CN=2$ chiral) multiplet ($\CE$). 
We find that the leading terms for the superconformal indices of rank 1 SCFTs  $H_1$, $H_2$, $D_4$, $E_6$, $E_7$ have the following universal expression in terms of the pletheystic exponential (PE): 
\begin{align} \label{eq:rk1Idx}
\begin{split}
 \CI_{\textrm{rank 1}} &= \PE \Big[ \CT + \bar\CE_{-\Delta} + \chi_{\textbf{adj}} \CJ  - \hat{\CB}_2 \chi_{\CI_2} \\
 &  \quad  - \bar{\CB}_{1, -2\Delta+1(0, 0)} 
  - \chi_{\textbf{adj}} \bar\CB_{1, -\Delta(0, 0)} - \bar{\CC}_{\half, -\Delta+\half (\half, 0)} - \chi_{\textbf{adj}}  \hat{\CC}_{1(0, 0)} + \ldots \Big] 
\end{split}
\end{align}
Here each symbol represents the indices for the corresponding short multiplets: for example, $\CT$ and $\CJ$ denote the indices for the stress-tensor multiplet and the conserved-current multiplet respectively. The $\bar{\CE}_{r}$ denotes the index for the $\CN=2$ chiral multiplet, which contains the Coulomb branch operator of dimension $\Delta = -r$. (See table \ref{table:shortmul}.)
And $\chi_{R}$ denotes the character of the representation $R$ of the flavor symmetry $\mathfrak{g}_F$. Here $\CI_2$ is given by the relation 
\begin{align}
 \textrm{Sym}^2 (\textbf{adj}) = (2 \cdot \textbf{adj}) \oplus \CI_2 \ , 
\end{align}
where $(2 \cdot \textbf{adj})$ refers to the representation with the Dynkin label being twice as those of the adjoint representation of the flavor symmetry $\mathfrak{g}_F$.
\begin{table}[t]
	\centering
	\begin{tabular}{c|ccccccc}
		 & $H_0$ & $H_1$ & $H_2$ & $D_4$ & $E_6$ & $E_7$ & $E_8$ \\
		 \hline
		 $\Delta$ & $\frac{6}{5}$ & $\frac{4}{3}$ & $\frac{3}{2}$ & $2$ & $3$ & $4$ & $6$ \\
		 $\mathfrak{g}_F$ & $\varnothing$ & $\mathfrak{su}_2$ & $\mathfrak{su}_3$ & $\mathfrak{so}_8$ & $\mathfrak{e}_6$ & $\mathfrak{e}_7$ & $\mathfrak{e}_8$ \\
		 $\textrm{dim}(\mathbf{adj})$ & $\cdot$ & ${3}$ & ${8}$ & ${28}$ & ${78}$ & ${133}$ & ${248}$ \\
		 $\textrm{dim}(\CI_2)$ & $\cdot$ & $1$ & $9$ & $106$ & $651$ &  $1540$ & $3876$ 
	\end{tabular}
	\caption{Data for the rank 1 $\CN=2$ `exceptional' superconformal theories. Here $\Delta$ is the dimension of the Coulomb branch operator and $\mathfrak{g}_F$ is the flavor symmetry. }
\end{table}
We are not able to check whether the expression \eqref{eq:rk1Idx} holds for the $E_8$ theory, since the full superconformal index for this theory is unknown. However, given the universality of this expression, we expect it to be true for the $E_8$ theory as well. 

From the full superconformal index and the selection rules for the OPE of short multiplets, we claim that the following OPE coefficients vanish for the rank 1 SCFTs $H_1, H_2, D_4, E_6, E_7, E_8$: 
\begin{align} \label{eq:vanishingope}
\begin{split}
\lambda [\CJ, \CJ, \hat\CB_2] \Big|_{\CI_2} &=0 \ , \\
\lambda [ \CE_{\Delta},  \CE_{\Delta} , {\CB}_{1, 2\Delta-1(0, 0)} ] &= 0\ ,  \\
\lambda [ \CE_{\Delta}, \CJ, \CB_{1, \Delta(0, 0)}]  &=0 \ , \\
\lambda [\CE_{\Delta}, \CT, \CC_{\half, \Delta-\half(0, \half)}] &=0 \ , \\
\lambda [\CT, \CJ, \hat\CC_{1(0, 0)}] &=0 \ , 
\end{split}
\end{align}
and also for their complex conjugates. The $\Delta$ denotes the dimension of the Coulomb branch operator. The first relation above is a well-known consequence of the Joseph relation for the 1-instanton moduli space. 

For the case of the $H_0$ theory, all the terms involving the conserved current multiplet $\CJ$ do not appear, so that we have a simpler form 
\begin{align}
 \CI_{H_0} = \textrm{PE} \left[ \CT + \bar{\CE}_{-\Delta} - \bar{\CB}_{1, -2\Delta+1(0, 0)} - \bar{\CC}_{\half, -\Delta+\half(\half, 0)} + \ldots \right] \ , 
\end{align}
with $\Delta=\frac{6}{5}$. It can be thought of as the same universal form as \eqref{eq:rk1Idx} upon removing all the flavor dependent terms. From the index and the OPE selection rules for $\CE \times \CT$ and $\CE \times \CE$, we find that the following OPE coefficients vanish in the $H_0$ theory:
\begin{align}
\begin{split}
 \lambda [\CE_\Delta, \CT, \CC_{\half, \Delta - \half(0, \half)} ] &=0 \ , \\
 \lambda [\CE_\Delta, \CE_\Delta, \CB_{1, 2\Delta - 1 (0, 0)} ] &=0 \ , 
\end{split}
\end{align} 
which is simply a subset of the relation \eqref{eq:vanishingope}. 

In the rest of the paper, we explain the procedure to determine the vanishing OPE coefficients from combining the superconformal index and the OPE selection rule. In section \ref{sec:4drank1}, we compute the full superconformal indices for the rank 1 SCFTs labeled by the Deligne-Cvitanovich exceptional series and find the vanishing short multiplets and show that our universal expressions \eqref{eq:rk1Idx} and \eqref{eq:vanishingope} hold. In section \ref{sec:5d}, we also apply our method to 5d $\CN=1$ SCFT and show that certain short multiplets vanish for the $E_1$ theory, very similar to that of the 4d $H_1$ theory. This hints that $E_1$ theory (UV fixed point of the $SU(2)_0$ SYM theory) is the direct 5d analog of the 4d $H_1$ Argyres-Douglas theory. 

\vspace{0.5cm}
\emph{Note added:} While the author was preparing the current paper, a very interesting paper \cite{Xie:2021omd} appeared on ArXiv, which has overlapping results with the current paper. The current paper covers additional vanishing short multiplets coming from the mixed OPEs, in addition to the rank 1 SCFTs that are not of Argyres-Douglas type. The paper of \cite{Xie:2021omd} covers higher-rank AD theories as well as their $\CN=1$ deformations. Part of the results in the current paper was announced at the APCTP workshop on Strings, Branes, and Gauge Theories 2020. 

\section{Vanishing short multiplets in rank 1 SCFTs} \label{sec:4drank1}

In this section, we compute the superconformal indices for the 4d $\CN=2$ rank 1 SCFTs. We will mostly use two different representations of the indices, one of which is given in terms of fugacities $(p, q, t)$ as
\begin{align}
 \CI(p, q, t) = \Tr (-1)^F p^{j_1+j_2+\frac{r}{2}} q^{j_2 - j_1 + \frac{r}{2}} t^{R-\frac{r}{2}} \ , 
\end{align}
where $j_{1, 2}$ denotes Cartans for the Lorentz group $SU(2)_1 \times SU(2)_2$ and $R, r$ denote the Cartans for the $SU(2)_R \times U(1)_r$. The index gets contributions from the states satisfying $\Delta \equiv E - 2j_2 - 2R - r/2 = 0$ and the fugacities satisfy
\begin{align}
 |p|<1, \quad |q|<1, \quad |t|<1, \quad \left| \frac{pq}{t} \right| < 1 \ . 
\end{align}
We sometimes use an alternative parametrization
\begin{align}
 p = \ft^3 y \ , \quad q=\ft^3/y \ , \quad t = \ft^4/v \ , 
\end{align}
with $|y|=|v|=1$ and $|\ft| < 1$, which is useful when expanding the index as a power series of $\ft$. Then the trace formula for the index can be written as
\begin{align}
 \CI(\ft, y, v) = \Tr (-1)^F \ft^{2(E + j_2)} y^{2j_1} v^{-R + \frac{r}{2}} \ .
\end{align}
One can take a simplification limit $p \to 0$ to obtain the Macdonald index, which counts the operators in the so-called Schur sector \cite{Gadde:2011uv}. One can further take $t \to q$ to obtain the Schur index, which is identical to the (vacuum) character of the associated vertex operator algebra (VOA) \cite{Beem:2013sza}. One can also take $p, q \to 0$ to obtain the Hall-Littlewood index or the Higgs branch index. For the rank one theories we consider, it computes the Hilbert series of the Higgs branch, which is identical to the (centered) one-instanton moduli space. It is given by a universal expression \cite{Benvenuti:2010pq, Keller:2011ek, Keller:2012da}
\begin{align}
 \CI_{HL} (t; \vec{a}) = \sum_n \chi_{(n \cdot \textbf{adj})}(\vec{a}) t^n \ ,  
\end{align}
where $(n \cdot \textbf{adj})$ denotes the irreducible representation labelled using the Dynkin label of adjoint representation multiplied by $n$.

\subsection{Indices and the selection rules for the short multiplets}
\begin{table}[t] 
	\centering
	\begin{tabular}{l | l | l}
	DO & CDI & comment \\
	\hline
	$\CC_{R, r(j_1, j_2)}$ & $A_\ell \bar{L}[j_1; j_2]^{(R;r)}$ & \\
	$\CB_{R, r(0, j)}$ & $B_1 \bar{L}[0; j]^{(R>0, r)}$ & \\
	$\CE_{r(0, j)}$ & $B_1 \bar{L}[0; j]^{(0; r)}$ & $\CE_r \equiv \CE_{r(0, 0)} $ \\
	$\hat{\CC}_{R(j_1, j_2)}$ & $A_\ell \bar{A}_{\bar{\ell}} [j_1; j_2]^{(R; j_1 - j_2)}$ & $\CT \equiv \hat{\CC}_{0(0, 0)}$ \\
	$\CD_{R(0, j)}$ & $B_1 \bar{A}_{\bar{\ell}}[0; j]^{(R; -j-2)}$ & \\
	$\hat{\CB}_R$ & $B_1 \bar{B}_1[0; 0]^{(R; 0)}$ & $\CJ \equiv \CB_1$  
	\end{tabular}
\caption{Short multiplets of 4d $\CN=2$ SCFTs in Dolan-Osborn (DO) and Cordova-Dumitrescu-Intriligator (CDI) notations.} 
\label{table:shortmul}
\end{table}

The superconformal index counts the degeneracies of short multiplets up to recombination \cite{Kinney:2005ej, Bhattacharya:2008zy}. Therefore for a given index, it is not always possible to extract the true short multiplet spectrum. On the other hand, we know the existence of certain universal multiplets such as stress-tensor ($\CT$), conserved-current ($\CJ$), Coulomb branch ($\CE_r$) and their OPE selection rules. Combining both information (sometimes with a plausible assumption), we are able to show that certain short multiplets are absent in the theory. 

The procedure we take is as follows \cite{Liendo:2015ofa, Song:2015wta, Agarwal:2018zqi}: Suppose we find the index of the form 
\begin{align}
 \CI = \PE \left[ \CI(A) + \CI(B) - \CI(C) + \ldots \right] \ ,
\end{align}
where $\CI(A), \CI(B), \CI(C)$ denotes the index for the short multiplets $A, B, C$, respectively (it may include a possibly minus sign). The multiplets $A$ or $B$ can be taken as universal ones such as stress-tensor, Coulomb branch, or conserved current. For this case, the contribution to the index is unambiguously coming from such universal multiplets. On the other hand, it is not uniquely specified which short multiplet contribute to $\CI(C)$ in general. This term may come from the vanishing of $C$ multiplet or come from another multiplet $C'$ that contributes to $\CI(C') = -\CI(C)$. 
However, if we once figure out the selection rules for the products of $A$ and $B$'s, such as $A \times A$, $A \times B$, $B \times B$ or products of powers of $A$ and $B$, one may look for the multiplet $C$ among these OPEs. Among the short multiplets that appear in these OPEs, if we find a unique multiplet $C$ that gives rise to $\CI(C)$, we can confidently claim that $C$ should disappear in the theory.\footnote{One may find such $C$ only at a multiple product of $A$ and $B$, for example $n$-fold product $A^n = A \times \cdots  \times A$. Indeed, such phenomenon happens for higher-rank theories \cite{Agarwal:2018zqi}. In such cases, one finds that the operator $C$ cannot appear in $A^i$ for $i < n$. } It does not completely rule out the possibility of having both multiplets $C$ and $C'$ so that we reproduce the index. However, it yields additional multiplets that contribute to higher-order with even larger cancellations. Therefore we claim the $C$ multiplet vanishes in this case, which is supported by the bootstrap results. 

Instead of computing the full selection rules for the OPEs, we take a short-cut by computing superconformal characters for the superconformal multiplets, say $\chi(A)$ and $\chi(B)$. Then we take the product $\chi(A) \times \chi(B)$ and decompose it into a sum over superconformal characters: $\chi(A) \times \chi(B) = \sum_{p} \chi(p)$ where $p$ collectively denotes the quantum numbers for the superconformal multiplets. This is \emph{not} the full selection rule for the OPE for the product of two local operators in $A$ and $B$ multiplets, since it misses contributions from singular pieces. Nevertheless, it turns out this is sufficient for us since what we need to find is an existence of certain multiplets in the OPE that can potentially contribute to the index. One may worry that singular contribution might give rise to additional short multiplet that might contribute to the same index, but this does not happen generically since it changes the scaling dimension. This does not fully justify our method, but we find our method agrees with the analysis based on OPE whenever the OPE for the supermultiplets are available.

\paragraph{Indices for the short multiplets}
Various short multiplets in 4d $\CN=2$ superconformal theories have been classified in \cite{Dolan:2002zh, Cordova:2016emh} (see table \ref{table:shortmul}) and their superconformal indices were computed in \cite{Gadde:2011uv}. Let us summarize the result here. 
\begin{itemize}
\item Index for the ${\CC}$ multiplet is given as
\begin{align}
	\CI_{ {\bar{\CC}}_{R, r(j_1, j_2) } } = (-1)^{2j_1 + 2j_2+1} t^{1+R+j_2} \left(\frac{pq}{t} \right)^{j_2-r} \frac{(1-\frac{pq}{t})(t-p)(t-q) \chi_{2j_1+1}\left(\sqrt{\frac{p}{q}} \right) }{(1-p)(1-q)}  , 
\end{align}	
where $\chi_{2j+1}(x)$ denotes $SU(2)$ character for the spin-$j$ representation, given as $\chi_{2j+1}(x) = \sum_{m=0}^{2j} x^{2j-m}$. 
\item Index for the $\hat{\CC}$ multiplet is given as 
\begin{align}
 \CI_{\hat{\CC}_{R(j_1, j_2)}} = (-1)^{2 j_1+2 j_2} t^{j_2+R+\frac{3}{2}}  \left(\frac{p q}{t}\right)^{j_1+\frac{1}{2}}  \frac{\left(1-\frac{p q}{t}\right) \left( \chi_{2 j_1+2}(\sqrt{\frac{p}{q}} )-\frac{\sqrt{p q} }{t} \chi_{2 j_1+1} (\sqrt{\frac{p}{q}}) \right) }{(1-p) (1-q)} . 
\end{align}
\item Index for the $\CB$ multiplet is given as
\begin{align}
 \CI_{\bar{\CB}_{R, r(j_1, 0)} }= \CI_{\bar\CC_{R-\half, r+\half(j_1, -\half)} } \ . 
\end{align}
\item Index for the $\hat{\CB}_R$ multiplet is given as
\begin{align}
 \CI_{\hat\CB_{R}} = \frac{t^R-p q t^{R-1}}{(1-p) (1-q)} \ . 
\end{align}

\item Index for the $\hat{\CE}_{r(0, 0)}$ multiplet is given as
\begin{align}
	\CI_{\bar{\CE}_{r(0, 0)} } = \left(\frac{p q}{t}\right)^{-r-1} \frac{(t-p) (t-q) }{t (1-p) (1-q)} \ . 
\end{align}
\end{itemize}

\paragraph{Selection rules for the OPE of short multiplets}
Let us summarize the selection rules for the OPE of universal short multiplets that we utilize in this paper. 
\begin{itemize}
\item OPE of two stress tensors \cite{Liendo:2015ofa} \cite{Agarwal:2018zqi}:
\begin{align}
 \hat{\CC}_{0(0, 0)} \times \hat{\CC}_{0(0, 0)} \sim \CI + \hat{\CC}_{0(\frac{\ell}{2}, \frac{\ell}{2})} + \hat{\CC}_{1(\frac{\ell}{2}, \frac{\ell}{2})} + \ldots
\end{align}
Here we sum over all possible positive integer $\ell$'s. 
\item OPE of chiral (Coulomb branch) multiplets \cite{Beem:2014zpa, Lemos:2015awa}:
\begin{align}
\begin{split}
 \CE_{r_1} \times \CE_{r_2} \sim \CE_{r_1+r_2} &+ \CB_{1, r_1+r_2-1(0, 0)} + \CC_{\half, r_1+r_2 - \frac{3}{2}(j - \half, j)} \\ &+ \CB_{\half, r_1 + r_2 - \half (0, \half)} + \CC_{0, r_1 + r_2 - 1 (j-1, j)} + \ldots
\end{split}
\end{align}
Here we sum over all half-integer and integer $j$'s. 
\item OPE of two conserved current multiplets:
\begin{align}
 \CJ \times \CJ = \hat{\CB}_1 \times \hat{\CB}_1 \sim \CI + \hat{\CB}_1 + \hat{\CB}_2 + \hat{\CC}_{0(j, j)} + \hat{\CC}_{1(j, j)} + \ldots
\end{align}
Here we sum over all half-integer and integer $j$'s. 
\item OPE of chiral multiplet and conserved current multiplet \cite{Gimenez-Grau:2020jrx}\footnote{There is an earlier work \cite{Ramirez:2016lyk} who worked out the OPE, but had an error which is fixed by \cite{Gimenez-Grau:2020jrx}. We also confirmed the result by computing superconformal characters.}:
\begin{align}
\CE_{r} \times \CJ &\sim \CB_{1, r(0, 0)} + \CC_{\half, r-\half(\frac{\ell-1}{2}, \frac{\ell}{2})} + \ldots
\end{align}
Here we sum over all possible positive integer $\ell$'s. 
\item The OPE of chiral and stress-tensor multiplet ($\CE \times \CT$) and the OPE of stress-tensor and conserved current multiplets ($\CT \times \CJ$) are worked out in \cite{Ramirez:2016lyk}, but our superconformal character computation does not seem to be compatible with the result of \cite{Ramirez:2016lyk}. Instead, we work out the leading order in the superconformal character to deduce the short-multiplets in these OPEs in the spirit of \cite{Agarwal:2018zqi}. 
\end{itemize}

\subsection{$H_0$ theory}
The full index of the $H_0$ theory can be computed using the $\CN=1$ gauge theory description given as $SU(2)$ gauge theory with 1 adjoint ($\phi$) and 2 fundamental ($q, q'$)  and 2 gauge-singlet ($M, X$) chiral multiplets with the superpotential \cite{Maruyoshi:2016tqk, Maruyoshi:2016aim, Fluder:2017oxm, Maruyoshi:2018nod}
\begin{align}
 W = X \phi^2 + M q' \phi q' +  \phi q q \ , 
\end{align}
where we omitted gauge indices. 
From this matter content and the interaction superpotential, one can obtain the contour integral expression for the superconformal index, which is given as
\begin{align}
 \CI_{H_0} = \frac{\G((\frac{pq}{t})^{\frac{6}{5}})}{\G((\frac{pq}{t})^{\frac{2}{5}})}  
  \frac{\kappa}{2} \oint_{|z|=1} \frac{dz}{2\pi i} \frac{\G(z^{\pm 2, 0} (\frac{p q}{t})^{\frac{1}{5}}) \G(z^{\pm} (\frac{pq}{t})^{\frac{2}{5}}t^{\half}) \G(z^{\pm} (\frac{pq}{t})^{-\frac{1}{5}}t^{\half})}{\G({z^{\pm 2})} } \ ,  
\end{align}
where we used the abbreviation $f(z^{\pm}) \equiv f(z) f(z^{-1})$ and $f(z^{\pm 2, 0}) \equiv f(z^2)f(z^{-2})f(z^0)$. Here $\kappa \equiv (p; p)(q; q)$ with $(z; q) = \prod_{n\ge 0} (1- z q^n)$ and $G(z)$ is the elliptic gamma function defined as
\begin{align}
 \G(z) \equiv \G (z; p, q) = \prod_{m, n \ge 0} \frac{1-z^{-1} p^{m+1} q^{n_+1}}{1-z p^m q^n} \ . 
\end{align}
As usual, we assume the fugacities to be $|p| < 1, |q|<1, |t|<1$ and $|pq/t|<1$. 

We find that the leading term of the resulting index can be rewritten in terms as
\begin{align} \label{eq:H0idx}
 \CI_{H_0} = \textrm{PE} \left[ \CT + \bar{\CE}_{-\frac{6}{5}} - \bar{\CB}_{1, -\frac{7}{5}(0, 0)} - \bar{\CC}_{\half, -\frac{7}{10}(\half, 0)}\right] + O(\ft^{\frac{51}{5}}) \ , 
\end{align}
where the expansion parameter $\ft$ is defined via the relation $p=\ft^3 y, q=\ft^3/y, t=\ft^4/v$. 
The product of the superconformal characters for the Coulomb branch and stress-tensor multiplet can be decomposed into
\begin{align} 
\begin{split}
 \chi_{\CE_{\frac{6}{5}}} \times \chi_\CT &\sim
\chi _{\CC_{0,\frac{6}{5}(0,0)}}+\chi _{\CC_{0,\frac{6}{5}(\frac{1}{2},\frac{1}{2})}}+\chi_{\CC_{0,\frac{6}{5}(1,1)}}+\chi _{\CC_{0,\frac{6}{5}(\frac{3}{2},\frac{3}{2})}}+\boxed{\chi_{\CC_{\frac{1}{2},\frac{7}{10}(0,\frac{1}{2})}} } + {\chi _{\CC_{\frac{1}{2},\frac{7}{10}(\frac{1}{2},0)}} } \\
&~~+\chi_{\CC_{\frac{1}{2},\frac{7}{10}(\frac{1}{2},1)}}+\chi _{\CC_{\frac{1}{2},\frac{7}{10}(1,\frac{1}{2})}}+\chi _{\CC_{\frac{1}{2},\frac{7}{10}(1,\frac{3}{2})}}+\chi _{\CC_{\frac{1}{2},\frac{7}{10}(\frac{3}{2},1)}}+\chi _{\CC_{\frac{1}{2},\frac{7}{10}(\frac{3}{2},2)}} + \cdots \ , 
\end{split}
\end{align}
where we omitted the long multiplets. The boxed short multiplet should appear in general in the OPE of Coulomb branch and stress tensor multiplets. However, the minus sign in front of $\bar\CC_{\half, -\frac{7}{10}(\half, 0)}$ inside the PE of the index \eqref{eq:H0idx} implies that such multiplet (and also its complex conjugate) should not appear in the theory. Had we just known the information from the index, one cannot discern which multiplet contributes to this particular term. But the OPE selection rule tells us that this is the unique choice. 

The product of the character for the two Coulomb branch operators decompose into
\begin{align}
\begin{split}
 \chi_{\CE_{\frac{6}{5}}} \times \chi_{\CE_{\frac{6}{5}}} &\sim 
 \chi _{\mathcal{E}_{\frac{12}{5}(0,0)}}+\chi _{\CB_{\frac{1}{2},\frac{19}{10}(0,\frac{1}{2})}}+\boxed{\chi _{\CB_{1,\frac{7}{5}(0,0)}}}+\chi _{\CC_{0,\frac{7}{5}(0,1)}}+\chi_{\CC_{0,\frac{7}{5}(\frac{1}{2},\frac{3}{2})}}+\chi _{\CC_{0,\frac{7}{5}(1,2)}} \\
 &~~+\chi _{\CC_{0,\frac{7}{5}(\frac{3}{2},\frac{5}{2})}}+\chi _{\CC_{\frac{1}{2},\frac{9}{10}(0,\frac{1}{2})}}+\chi _{\CC_{\frac{1}{2},\frac{9}{10}(\frac{1}{2},1)}}+\chi _{\CC_{\frac{1}{2},\frac{9}{10}(1,\frac{3}{2})}}+\chi_{\CC_{\frac{1}{2},\frac{9}{10}(\frac{3}{2},2)}} + \cdots \ . 
\end{split}
\end{align}
Combining the information from the index \eqref{eq:H0idx}, we see that the boxed operator above is the one that are absent in the $H_0$ theory. Therefore the vanishing OPE coefficients for the $H_0$ theory are
\begin{align}
 \lambda \left[\CE_{\frac{6}{5}}, \CT, \CC_{\half, \frac{7}{10} (\half, 0)} \right] &=0 \ ,\\
 \lambda \left[\CE_{\frac{6}{5}}, \CE_{\frac{6}{5}}, \CB_{1, \frac{7}{5} (0, 0)} \right] &=0 \ . 
\end{align}
It is also known that the following OPE vanishes:
\begin{align}
 \lambda \left[ \CT, \CT, \hat{\CC}_{1(\half, \half)} \right] &=0 \ .
\end{align}
It can be obtained from the Macdonald index \cite{Song:2015wta} and is the condition for the central charge $c$ to saturate the bound $c \ge \frac{11}{30}$ \cite{Liendo:2015ofa}. 

\subsection{$H_1$ theory}
The full index for the $H_1$ theory can be obtained by considering $\CN=1$ gauge theory description given as the $SU(2)$ gauge theory with 1 adjoint ($\phi$) and 2 fundamentals $(q, q')$ and 2 gauge-singlets $(M, X)$ with the superpotential \cite{Maruyoshi:2016aim, Agarwal:2016pjo, Agarwal:2017roi}\footnote{This theory has two dual $\CN=1$ Lagrangian descriptions coming from the fact $H_1=(A_1, A_3)=(A_1, D_3)$. We utilize the $(A_1, A_3)$ description here.}
\begin{align}
 W = X  \phi^2 + M q q' \ .
\end{align}
The full superconformal index can be obtained via evaluating the following integral:
\begin{align}
 \CI_{H_1} = \frac{\G((\frac{pq}{t})^{\frac{4}{3}})}{\G((\frac{pq}{t})^{\frac{2}{3}})}  
  \frac{\kappa}{2} \oint_{|z|=1} \frac{dz}{2\pi i} \frac{\G(z^{\pm 2, 0} (\frac{p q}{t})^{\frac{1}{3}}) \G(z^{\pm} a (\frac{pq}{t})^{-\frac{1}{6}}t^{\half}) \G(z^{\pm} a^{-1} (\frac{pq}{t})^{-\frac{1}{6}}t^{\half})}{\G({z^{\pm 2})} } \ .
\end{align}
Here $a$ is the fugacity for the $SU(2)$ flavor symmetry. We find that the above index can be written in terms of plethystic exponential as
\begin{align}
\begin{split}
 \CI_{H_1} &= \PE \Big[ \chi_{\textrm{adj}} \CJ + \bar\CE_{-\frac{4}{3}} + \CT - \hat{\CB}_2 - \bar{\CB}_{1, -\frac{5}{3}(0, 0)} \\ 
  & \qquad \qquad - \chi_{\textrm{adj}} \bar\CB_{1, -\frac{4}{3}(0, 0)} - \bar{\CC}_{\half, -\frac{5}{6} (\half, 0)} -  \hat{\CC}_{1(0, 0)} \chi_{\textrm{adj}} + O(\ft^{10}) \Big] \ . 
\end{split}
\end{align}
One might wonder why we have the stress-tensor ($\CT$) as a separate generator since it can appear in the OPE of $\CJ \times \CJ$. Indeed, in the associated VOA, the (2d) stress tensor is given by the Sugawara construction so that there is no independent $\CT$. However, we find that the index cannot be written in terms of PE without separate $\CT$ contribution. This can be already seen by taking the Macdonald limit $p \to 0$, which yields
\begin{align}
 \CI_{H_1} = \PE \left[ \frac{q T \chi_{\textrm{adj}} + q^2 (T-T^2) - q^3 T^2 \chi_{\textrm{adj}}}{1-q} + O(q^4) \right] \ , 
\end{align}
where $t=q T$. The $q T$ and $q^2 T$ terms come from the conserved current and stress-tensor multiplets respectively. The terms $q^2 T^2$ and $q^3 T^2$ come from $\hat\CB_2$ and $\hat\CC_{1(0, 0)}$ respectively. Notice that only after taking the Schur limit $T \to 1$, the $q^2$ term vanishes. 

We obtain the following selection rules involving mixed OPE of Coulomb branch operator and stress tensor or conserved current from the superconformal character decomposition: 
\begin{align}
\begin{split}
 \chi_{\CE_{\frac{4}{3}} } \times \chi_{\CT} & \sim \chi_{\CC_{0,\frac{4}{3}(0,0)}} + \boxed{\chi _{\CC_{\frac{1}{2},\frac{5}{6}(0,\frac{1}{2})}} }+\chi _{\CC_{0,\frac{4}{3}(\frac{1}{2},\frac{1}{2})}}  + \chi _{\CC_{\frac{1}{2},\frac{5}{6}(\frac{1}{2},1)}} + \chi _{\CC_{\frac{1}{2},\frac{5}{6}(\frac{1}{2},0)}} + \chi _{\CC_{0,\frac{4}{3}(1,1)}} \\ 
 &\qquad + \chi _{\CC_{\frac{1}{2},\frac{5}{6}(1,\frac{3}{2})}} + \chi _{\CC_{\frac{1}{2},\frac{5}{6}(1,\frac{1}{2})}} + \chi _{\CC_{0,\frac{4}{3} (\frac{3}{2},\frac{3}{2})}} + \chi _{\CC_{\frac{1}{2},\frac{5}{6}(\frac{3}{2},2)}} + \chi _{\CC_{\frac{1}{2},\frac{5}{6}(\frac{3}{2},1)}} + \ldots
\end{split} \\
\begin{split}
 \chi_{\CE_{\frac{4}{3}}} \times \chi_\CJ &\sim \boxed{\chi _{\CB_{1,\frac{4}{3}(0,0)}}} + \chi _{\CC_{\frac{1}{2},\frac{5}{6}(0,\frac{1}{2})}}  + \chi _{\CC_{\frac{1}{2},\frac{5}{6}(\frac{1}{2},1)}} + \chi _{\CC_{\frac{1}{2},\frac{5}{6}(1,\frac{3}{2})}}+ \chi _{\CC_{\frac{1}{2},\frac{5}{6}(\frac{3}{2},2)}} + \ldots
\end{split}\\
\begin{split}
\chi_{\CT} \times \chi_{\CJ} & \sim
\boxed{\chi _{\hat{\CC}_{1(0,0)}} } +\chi _{\hat{\CC}_{1(\frac{1}{2},\frac{1}{2})}}+\chi _{\hat{\CC}_{1(1,1)}}+\chi _{\hat{\CC}_{1(\frac{3}{2},\frac{3}{2})}}+\chi _{\hat{\CC}_{1(2,2)}}+\chi _{\CC_{\frac{1}{2},\frac{1}{2}(\frac{1}{2},0)}} + \ldots
\end{split}
\end{align}
The boxed short multiplets are the ones that are vanishing. Here we also need to take the flavor part of the index into account, in order to obtain the correct vanishing condition. 

The relevant OPE selection rules for us are given as 
\begin{align}
\begin{split}
 \hat{\CB}_2 &\in \CJ \times \CJ \ , \\
 \bar{\CB}_{1, -\frac{5}{3}(0, 0)} &\in \bar{\CE}_{-\frac{4}{3}} \times \bar{\CE}_{-\frac{4}{3}} \ , \\
 \bar\CB_{1, -\frac{4}{3}(0, 0)} &\in \bar{\CE}_{-\frac{4}{3}} \times \CJ \ , \\
 \bar{\CC}_{\half, -\frac{5}{6} (\half, 0)} &\in \bar{\CE}_{-\frac{4}{3}} \times \CT \ , \\
  \hat\CC_{1(0, 0)} &\in \CT \times \CJ \ , 
\end{split}
\end{align}
and also their complex conjugates. 
From the expression for the index and the selection rule above, we find that the following OPE coefficients vanish in the $H_1$ theory:
\begin{align}
\begin{split}
   \lambda \left[ \CJ, \CJ, \hat\CB_2 \right] \Big|_{\CI_2} &= 0\\
 \lambda \left[\CE_{\frac{4}{3}}, \CE_{\frac{4}{3}}, \CB_{1, \frac{5}{3}(0, 0)} \right]&=0 \\
  \lambda \left[\CE_{\frac{4}{3}}, \CJ, \CB_{1, \frac{4}{3}(0, 0)} \right]&=0 \\
   \lambda \left[\CE_{\frac{4}{3}}, \CJ, \CC_{1, \frac{5}{6}(0, \half)}\right] &=0 \\
   \lambda \left[ \CT, \CJ, \hat\CC_{1(0, 0)} \right] &= 0
\end{split}
\end{align}
For the $H_1$ theory with $SU(2)$ flavor symmetry, $\CI_2$ simply consists of the flavor singlet since $\mathrm{Sym}^2(\mathbf{adj}) = \mathbf{5} \oplus \mathbf{1}$ in $SU(2)$. 

\subsection{$H_2$ theory}
We can compute the full index for the $H_2$ theory using the $\CN=1$ Lagrangian description given by $SU(2)$ gauge theory with 1 adjoint $(\phi)$ and 4 fundamentals $(q_1, \tilde{q}_1, q_2, \tilde{q}_2)$ and 2 singlets $(M, X)$ with the superpotential \cite{Agarwal:2016pjo, Agarwal:2017roi}:
\begin{align}
 W = X \phi^2 + M q_1 \tilde{q}_1 + q_2 \phi \tilde{q}_2 
\end{align}
The Lagrangian only has $SU(2) \times U(1)$ as its manifest flavor symmetry, but it enhances to $SU(3)$ in the IR upon RG flow. 
Now, the index can be computed as
\begin{align}
 \CI_{H_2} = \frac{\G((\frac{pq}{t})^{\frac{3}{2}})}{\G((\frac{pq}{t}))}  
  \frac{\kappa}{2} \oint_{|z|=1} \frac{dz}{2\pi i} \frac{\G(z^{\pm 2, 0} (\frac{p q}{t})^{\frac{1}{2}}) \prod_{i=\pm} \G(z^{\pm} x_i (\frac{pq}{t})^{\frac{1}{4}}t^{\half}) \G(z^{\pm} y_i (\frac{pq}{t})^{-\frac{1}{4}}t^{\half})}{\G({z^{\pm 2})} } \ ,
\end{align}
where $x_{\pm} = (a_1 a_2^3)^{\pm 1}$, $y_{\pm} = (a_1 a_2^{-1})^{\pm 1}$ with $a_1, a_2$ being $SU(3)$ flavor fugacities. 
Upon evaluating the index, we find the following expression: 
\begin{align}
\begin{split}
 \CI_{H_2} &= \PE \Big[ \chi_{\textrm{adj}}(a) \CJ + \bar\CE_{-\frac{3}{2}} + \CT - \hat{\CB}_2 (1+\chi_{\textrm{adj}}(a)) - \bar{\CB}_{1, -2(0, 0)}\\ 
  & \qquad \qquad  - \chi_{\textrm{adj}}(a) \bar\CB_{1, -\frac{3}{2}(0, 0)} - \bar{\CC}_{\half, -1(\half, 0)} - \hat{\CC}_{1(0, 0)}  \chi_{\textrm{adj}}(a) + O(\ft^{11}) \Big]
\end{split}
\end{align}
Repeating the same procedure, we find the relevant OPEs to be
The relevant OPE selection rules for us are given as 
\begin{align}
\begin{split}
 \hat{\CB}_2 &\in \CJ \times \CJ \ , \\
 \bar{\CB}_{1, -2(0, 0)} &\in \bar{\CE}_{-\frac{3}{2}} \times \bar{\CE}_{-\frac{3}{2}} \ , \\
 \bar\CB_{1, -\frac{3}{2}(0, 0)} &\in \bar{\CE}_{-\frac{3}{2}} \times \CJ \ , \\
 \bar{\CC}_{\half, -1 (\half, 0)} &\in \bar{\CE}_{-\frac{3}{2}} \times \CT \ , \\
  \hat\CC_{1(0, 0)} &\in \CT \times \CJ \ , 
\end{split}
\end{align}
and their complex conjugates as well. Combining the information from the index and the selection rule, we find the following OPE coefficients (and their complex conjugates) vanish in the $H_2$ theory:
\begin{align}
\begin{split}
   \lambda \left[ \CJ, \CJ, \CB_2 \right] \Big|_{\CI_2} &= 0\\
 \lambda \left[\CE_{\frac{3}{2}}, \CE_{\frac{3}{2}}, \CB_{1, 2(0, 0)} \right]&=0 \\
  \lambda \left[\CE_{\frac{3}{2}}, \CJ, \CB_{1, \frac{3}{2}(0, 0)} \right]&=0 \\
   \lambda \left[\CE_{\frac{3}{2}}, \CJ, \CC_{\half, 1(0, \half)}\right] &=0 \\
   \lambda \left[ \CT, \CJ, \hat\CC_{1(0, 0)} \right] &= 0
\end{split}
\end{align}
Here $\CI_2 = \mathbf{8} \oplus \mathbf{1}$ for $SU(3)$. 

\subsection{$D_4$ theory}
The $D_4$ theory is a Lagrangian superconformal gauge theory of $SU(2)$ with 4 fundamental hypermultiplets. The index can be computed using the following integral:
\begin{align}
 \CI_{D_4} &= \frac{\kappa}{2} \oint_{|z|=1} \frac{dz}{2\pi i} \frac{\Gamma \left(\frac{pq}{t} z^{\pm 2, 0} \right) \prod_{i=1}^4 \Gamma(z^{\pm} a_i t^\half)\Gamma(z^{\pm} a_i^{-1} t^\half) }{\Gamma(z^{\pm 2})}
\end{align}
Here $a_{i=1,2, 3,4}$ are the fugacities for the $SO(8)$ flavor symmetry. 
We find that the index can be written as
\begin{align}
\begin{split}
 \CI_{D_4} &= \PE \Big[ \bar \CE_{-2} + \CT + \CJ \chi_{\textrm{adj}} - \hat\CB_{2} \chi_{\CI_2} - \bar\CB_{1, -2(0, 0)} \chi_{\textrm{adj}}(\vec{a}) \\
 & \qquad \qquad - \bar\CB_{1, -3(0, 0)} - \hat{\CC}_{1(0, 0)} \chi_{\textrm{adj}} - \bar\CC_{\half, -\frac{3}{2}(\half, 0)} + O(\ft^{12}) \Big]
\end{split}
\end{align}
where $\CI_2 = \textbf{35}_v \oplus \textbf{35}_c \oplus \textbf{35}_s \oplus \textbf{1}$. 
The relevant OPE selection rules for us are given as 
\begin{align}
\begin{split}
 \hat{\CB}_2 &\in \CJ \times \CJ \ , \\
 \bar{\CB}_{1, -3(0, 0)} &\in \bar{\CE}_{-2} \times \bar{\CE}_{-2} \ , \\
 \bar\CB_{1, -2(0, 0)} &\in \bar{\CE}_{-2} \times \CJ \ , \\
 \bar{\CC}_{\half, -\frac{3}{2} (\half, 0)} &\in \bar{\CE}_{-2} \times \CT \ , \\
  \hat\CC_{1(0, 0)} &\in \CT \times \CJ \ , 
\end{split}
\end{align}
and their complex conjugates as well. From the index and the selection rules, we find that the following OPE coefficients vanish:
\begin{align}
\begin{split}
   \lambda \left[ \CJ, \CJ, \hat\CB_2 \right] \Big|_{\CI_2} &= 0\\
 \lambda \left[\CE_{2}, \CE_{2}, \CB_{1, 3(0, 0)} \right]&=0 \\
  \lambda \left[\CE_{2}, \CJ, \CB_{1, 2(0, 0)} \right]&=0 \\
   \lambda \left[\CE_{2}, \CJ, \CC_{\half, \frac{3}{2}(0, \half)}\right] &=0 \\
   \lambda \left[ \CT, \CJ, \hat\CC_{1(0, 0)} \right] &= 0
\end{split}
\end{align}

\subsection{$E_6$ theory}
The full index is computed in \cite{Gadde:2010te} utilizing Argyres-Seiberg duality between partially gauged $E_6$ theory and $SU(3)$ $N_f=6$ SQCD. It can be also understood as an $\CN=1$ (singular) Lagrangian description for the $E_6$ theory \cite{Gadde:2015xta}. This yields the integral expression for the index to be
\begin{align}
 \CI_{E_6} = \frac{\kappa}{2} \Gamma(t w^{\pm 2}) \Gamma(pqt) \oint_{C_w} \frac{ds}{2\pi i s} \frac{\Gamma(t^{-\half} s^{\pm 1} w^{\pm 1})}{ \Gamma(s^{\pm 2})} \CI_{SU(3)}(s, \vec{a}) \ , 
\end{align}
where the integration contour $C_w$ is almost a unit circle but deformed so that it includes the poles at $s=w^{\pm 1} t^{-\half}$ but not the ones at $s=w^{\pm 1} t^{\half}$. 
The index for the $SU(N)$ theory is given as
\begin{align} \label{eq:idxSUN}
 \CI_{SU(N)} (s, \vec{a}) = \frac{(\kappa \Gamma(\frac{pq}{t}))^{N-1}}{N!} \oint \prod_{i=1}^N \frac{d z_i}{2\pi i} \frac{\prod_{i=1}^N \prod_{f=1}^{2N} \Gamma(t^{\half} (s^{\frac{1}{N}} z_i a_f)^{\pm 1}) \prod_{i \neq j} \Gamma (\frac{pq}{t} \frac{z_i}{z_j})}{\prod_{i \neq j} \Gamma(\frac{z_i}{z_j})} \ . 
\end{align}
Upon evaluating the integral, we find the index for the $E_6$ Minahan-Nemeschansky theory can be written in terms of PE as 
\begin{align}
\begin{split}
 \CI_{E_6} &= \PE \Big[ \bar \CE_{-3} + \CT + \CJ \chi_{\textrm{adj}} - \hat\CB_2 \chi_{\CI_2}  - \bar\CB_{1, -3(0, 0)} \chi_{\textrm{adj}} - \hat{\CC}_{1(0, 0)} \chi_{\textrm{adj}}  \\
 & \qquad \qquad  - \bar{\CC}_{\half, -\frac{5}{2}(\half, 0)}  + 12376 \hat\CB_3 + \frac{\ft^{13}}{v} \chi_2(y) - \bar{\CB}_{1, -5(0, 0)} + O(\ft^{14}) \Big] \ ,
\end{split}
\end{align}
where $\CI_2 = \textbf{650} \oplus \textbf{1}$ and $\chi_2(y) = y+1/y$ is the character for the angular momentum. The term proportional to $\hat\CB_3$ is present so that the Hall-Littlewood limit of the index at order $t^3$ has the coefficient $\chi_{3\cdot \mathbf{adj}}$. For example, the Hall-Littlewood index for the $E_7$ theory can be written as 
\begin{align}
\begin{split}
 \CI_{E6}^{HL} (t) &= \sum_{n} \chi_{n \cdot \textbf{adj}} t^n = 1 + 78 t + 2430 t^2 + 43758 t^3 + O(t^4) \\
 &\qquad =  \PE \left[ 78 t -651 t^2 + 12376 t^3 + \ldots \right] \ . 
\end{split}
\end{align}
The extra term present at this order is due to the fact that $\bar{\CC}_{\half, -\frac{5}{2}(\half, 0)}$ is of order $\ft^{13}$. The index for the $\bar\CB_{1, -5(0, 0)}$ is of $O(\ft^{14})$ and we indeed find that our expression is consistent with the universal form \eqref{eq:rk1Idx}. But there exists another terms at this order that we do skip identifying with particular short multiplets. It does not alter our general claim of the vanishing OPE coefficients. We again find the following OPE coefficients vanish:
\begin{align}
\begin{split}
   \lambda \left[ \CJ, \CJ, \hat\CB_2 \right] \Big|_{\CI_2} &= 0\\
 \lambda \left[\CE_{3}, \CE_{3}, \CB_{1, 5(0, 0)} \right]&=0 \\
  \lambda \left[\CE_{3}, \CJ, \CB_{1, 3(0, 0)} \right]&=0 \\
   \lambda \left[\CE_{3}, \CJ, \CC_{\half, \frac{5}{2}(0, \half)}\right] &=0 \\
   \lambda \left[ \CT, \CJ, \hat\CC_{1(0, 0)} \right] &= 0
\end{split}
\end{align}

\subsection{$E_7$ theory}
The full index is computed in \cite{Agarwal:2018ejn} utilizing an $\CN=1$ Lagrangian description, which generalizes the $E_6$ theory case by considering the S-duality of $SU(4)$ $N_f=8$ SQCD \cite{Chacaltana:2010ks} and a partial Higgsing. This yields the integral expression for the index to be
\begin{align}
 \CI_{E_7} = \frac{\kappa}{2} \Gamma(t^2) \Gamma(pqt) \Gamma \left(\frac{pq}{t} \right) \oint_{C_s} \frac{ds}{2\pi i s} \frac{\Gamma(s^{\pm 1}) \Gamma(t^{-1} s^{\pm 1})}{\Gamma(s^{\pm 2})} \CI_{SU(4)} (s) \ , 
\end{align}
where the integration contour $C_s$ is almost a unit circle but deformed in such a way to include the poles at $s=t^{-1}$ but not $s=t^1$. The index $\CI_{SU(4)}$ for the $SU(4)$ SQCD is given in \eqref{eq:idxSUN}. Upon evaluating the integral, we find the leading order expression can be written as
\begin{align}
\begin{split}
  \CI_{E_7} &= \PE \Big[ \bar\CE_{-4} + \CT + \CJ \chi_{\textrm{adj}} - \hat\CB_2 (\chi_{\textbf{1539}} + \chi_{\textbf{1}}) - \hat\CC_{1(0, 0)} \chi_{\textrm{adj}} \\ 
  &\qquad \qquad - \bar\CB_{1, -4(0, 0)} \chi_{\textrm{adj}} + 42427 \hat\CB_3  - \bar{\CC}_{\half, -\frac{7}{2}(\half, 0)} - \bar\CB_{1, -5(0, 0)} + O(\ft^{13})  \Big] \ . 
\end{split}
\end{align}
Once again the coefficient in front of $\hat{\CB}_3$ makes the Hall-Littlewood limit of the index at order $t^3$ to be of the form $\chi_{n \cdot \textrm{adj}}$: 
\begin{align}
\begin{split}
 \CI_{E7}^{HL} (t) &= \sum_{n} \chi_{n \cdot \textbf{adj}} t^n = 1 + 133 t + 7371 t^2 + 238602 t^3 + O(t^4) \\
 &\qquad =  \PE \left[ 133 t - 1540 t^2 + 42427 t^3 + \ldots \right] \ . 
\end{split}
\end{align}
The index for the short multiplets $\bar{\CC}_{\half, -\frac{7}{2}(\half, 0)}$ and $\bar\CB_{1, -5(0, 0)} $ is of order $O(\ft^{15})$ and $O(\ft^{18})$, respectively. Therefore we have many more terms contributing to the index at this order. However, we do find that the universal form of the index is consistent with \eqref{eq:rk1Idx} apart from superfluous terms coming from other short multiplets that we do not identify with. Therefore, we find the following OPE coefficients vanish:
\begin{align}
\begin{split}
   \lambda \left[ \CJ, \CJ, \hat\CB_2 \right] \Big|_{\CI_2} &= 0\\
 \lambda \left[\CE_{4}, \CE_{4}, \CB_{1, 7(0, 0)} \right]&=0 \\
  \lambda \left[\CE_{4}, \CJ, \CB_{1, 4(0, 0)} \right]&=0 \\
   \lambda \left[\CE_{4}, \CJ, \CC_{\half, \frac{7}{2}(0, \half)}\right] &=0 \\
   \lambda \left[ \CT, \CJ, \hat\CC_{1(0, 0)} \right] &= 0
\end{split}
\end{align}
Given the universality of the index so far, we claim that the form of the full index \eqref{eq:rk1Idx} and the vanishing OPE \eqref{eq:vanishingope} to be true for the $E_8$ theory, even though we are not able to explicitly compute the full index. It is tempting to ask if this relation is true for other rank one SCFTs that do not realize one instanton moduli space as its Higgs branch. They do not satisfy the condition $\lambda \left[ \CJ, \CJ, \hat\CB_2 \right] \Big|_{\CI_2} = 0$, but it still remains to be seen if any other relations hold in such cases as well. 

%%%%%%%%%%%%%%%%%%%%%%%%%%%%%%%%%%%%%%%%%
\section{Vanishing short multiplets in 5d $E_1$ theory} \label{sec:5d}
\subsection{Superconformal index in 5d}
Superconformal index for five-dimensional $\CN=1$ superconformal theory was first defined in \cite{Bhattacharya:2008zy} and computed for a number of UV fixed points of supersymmetric gauge theories using localization in \cite{Kim:2012gu}. It is defined as
\begin{align}
 \CI(x, y) = \tr (-1)^F  x^{2j_1 + R} y^{2j_2} \prod_i {a_i}^{F_i} \ , 
\end{align}
where $j_1, j_2$ are the Cartans for the Lorentz group $SU(1)_1 \times SU(2)_2 \subset Sp(2)$ and $R$ being the Cartan of the $SU(2)_R$ symmetry. The $F_i$'s denotes the Cartans for the flavor symmetry group. The trace gets non-vanishing contributions only from the states with $\Delta \equiv E - 2 j_1 - 3 R = 0$. 

Short multiplets in 5d SCFT are classified by Cordova-Dumitrescu-Intriligator (CDI) \cite{Cordova:2016emh} and also by Buican-Hayling-Papageorgakis (BHP) \cite{Buican:2016hpb}. The superconformal index for the conserved current ($\CJ = C_1[0, 0]_3^{(2)}$ in the CDI notation and $\CJ = \CD[0, 0; 2]$ in the BHP notation) and the stress-tensor ($\CT = B_2[0, 0]_3^{(0)}$ in CDI and $\CT = \CB[0, 0;0]$ in BHP) multiplets are given as \cite{Buican:2016hpb}
\begin{align}
 \CI_\CJ = \CI_{\CD[0, 0;2]} = \frac{x^2}{(1-xy)(1-x/y)} \ , \quad \CI_\CT =\CI_{\CB[0, 0;0]} = \frac{x^3 \chi_{\mathbf{2}}(y)}{(1-xy)(1-x/y)} \ , 
\end{align}
where $\chi_{\mathbf{n}}(y)$ denotes the character for the $n$-dimensional representation of $SU(2)$. The denominator comes from the conformal descendants generated by derivatives. We summarize the indices for the short multiplets below. 
\begin{itemize}
	\item The $\CA$-type (BHP)/$A$-type (CDI) multiplets: 
		\begin{align}
			\CI_{\CA[d_1, d_2; R]} = (-1)^{d_2+1} \frac{x^{d_1+d_2+R+4}}{(1-xy)(1-x/y)} \chi_{d_1+1}(y)
		\end{align}
		Here $d_1, d_2$ are the Dynkin labels for the $SU(2)$ representations, which are twice as that of the spins $(d_i = 2j_i)$. 
	\item The $\CB$-type (BHP)/$B$-type (CDI) multiplets:
		\begin{align}
			\CI_{\CB[d_1, 0; R]} = \frac{x^{d_1+R+3}}{(1-xy)(1-x/y)} \chi_{d_1+2}(y)
		\end{align}
	\item The $\CD$-type (BHP)/$C$-type (CDI) multipets:
		\begin{align}
			\CI_{\CD[0, 0; R]} = \frac{x^R}{(1-xy)(1-x/y)}  
		\end{align}
\end{itemize}
We can follow the same procedure as in the case of 4d theory, namely combining the information from the index and the selection rules to deduce the vanishing OPE coefficients. As far as the author is aware, unlike the case of 4d $\CN=2$ theories, the (mixed) OPEs in 5d superconformal theories are still not carried out in detail. (See  \cite{Chang:2017cdx, Bobev:2017jhk, Baume:2019aid} for the work on 5d superconformal blocks and bootstrap.) Here we resort to a simple group theory argument, which is sufficient for our purpose. 

\subsection{$E_1$ theory} 
The five-dimensional rank 1 superconformal theories are discovered by Seiberg \cite{Seiberg:1996bd}, which are realized as UV fixed points of $SU(2)$ gauge theory of $N_f \le 7$ fundamental hypermultiplets. They are known to exhibit enhanced global symmetry of $E_{N_f+1}$ and such an enhancement of symmetry is verified using the superconformal index \cite{Kim:2012gu, Bashkirov:2012re, Hwang:2014uwa} and via study of the instanton operators \cite{Tachikawa:2015mha}. 

Here, let us focus on the minimal case of $N_f=0$. The leading terms in the superconformal index of the $E_1$ theory can be written as
\begin{align} \label{eq:E1idx}
 \CI_{E_1} = \PE \left[ \frac{\chi_{\mathbf{3}} (a) x^2 + \chi_{\mathbf{2}}(y) x^3 - x^4 - \chi_{\mathbf{3}}(a) \chi_{\mathbf{2}}(y) x^5 - \left(\chi_{\mathbf{3}}(a) + 1 + \chi_{\mathbf{3}}(y) \right) x^6 + \ldots }{(1-xy)(1-x/y)}\right] \ . 
\end{align}
We see that the first two terms of the index comes from the conserved current ($x^2$) and the stress-tensor ($x^3$) multiplets. From the $x^4, x^5, x^6$ terms, we see that certain short multiplets that appear in $\CJ \times \CJ$, $\CJ \times \CT$ and $\CT \times \CT$ should disappear. 

The $-x^4$ term in \eqref{eq:E1idx} can be reproduced by the absence of the $\CD[0, 0;4]$ multiplet which is generally present in the OPE of $\CJ \times \CJ = \CD[0, 0; 2] \times \CD[0, 0; 2]$. Therefore, we have the following OPE coefficient vanish:
\begin{align}
 \CD[0, 0;4] \in \CJ \times \CJ \  \longrightarrow \ \lambda\left[ \CJ, \CJ, \CD[0, 0;4] \right] \Big|_{\CI_2} = 0 \ , 
\end{align}
where $\CI_2 = \mathbf{1}$ is the singlet of the $SU(2)$ flavor. 

The $-x^5 \chi_{\mathbf{3}}(a) \chi_{\mathbf{2}}(y)$ term can be obtained either by the absence of the $\CB[0, 0;2]$ multiplet or an extra $SU(2)$ triplet of $\CA[1, 0;0]$ multiplet. But we find that $\CB[0, 0;2]$ multiplet exists in the OPE of $\CT \times \CJ = \CB[0, 0;0] \times \CD[0, 0; 2] $ but not $\CA[1, 0;0]$, the short multiplet $\CB[0, 0;2]$ should vanish in this theory. Hence, we have
\begin{align}
\begin{split}
 \CB[0, 0;2] \in \CT \times \CJ  , ~~ \CA[1, 0;0] \notin \CT \times \CJ \, 
  \longrightarrow \ \lambda\left[ \CT, \CJ, \CB[0, 0; 2] \right]  = 0 \ . 
\end{split}
\end{align}

The terms at order $x^6$ has two parts: One from $\CJ^3$ and the other from $\CT^2$. The $\CJ^3 = \CD[0,0;2]^3$ part arises from $\CD[0, 0;4] \times \CD[0, 0;4]$ which includes $\CD[0, 0;6]$, and this is the only short multiplet that gives rise to $- \chi_{\mathbf{3}}(a) x^6$. Therefore, the following OPE coefficients should vanish
\begin{align}
\begin{split}
 \CD[0, 0; 6] \in \CJ^3 \ 
 \longrightarrow \ \lambda\left[\CJ, \CD[0, 0;4], \CD[0, 0; 6] \right] \Big|_{\textrm{adj}} = 0 \ . 
\end{split}
\end{align}
The term $-\chi_{\mathbf{3}}(y) x^6$ can be obtained from two short multiplets: vanishing of $\CB[1, 0;2]$ or addition of $\CA[2, 0; 0]$ multiplet. The former appears in the OPE of $\CT \times \CT$, but the latter does not. Therefore, we find the vanishing OPE
\begin{align}
\CB[1, 0;2] \in \CT \times \CT , ~~ \CA[2, 0;0] \notin \CT \times \CT \, \longrightarrow \,
 \lambda \left[ \CT, \CT, \CB[1, 0;2] \right] = 0 \ . 
\end{align}
The singlet term $-x^6$ can come from the vanishing of $\CA[0, 1;1]$ or addition of $\CA[0, 0;2]$ or addition of $\CA[0, 2;0]$. However among them only $\CA[0, 1;1]$ multiplet can appear in $\CT \times \CT$. Therefore, we have
\begin{align}
 \CA[0, 1;1] \in \CT \times \CT \ \longrightarrow \ \lambda \left[\CT, \CT, \CA[0, 1;1] \right] = 0 \ . 
\end{align}

The vanishing structure is quite similar to that of the 4d rank one theories we considered. It would be interesting to see whether we have some kind of universality for other 5d rank one SCFTs. 
Let us notice that we also have $\CJ^3 \big|_{\textrm{adj}} \sim 0$, which is identical to that of the $H_1$ theory in 4d. In fact, a similar relation $\CJ^{2n+1}\big|_{\textrm{adj}} \sim 0$ holds for all the $(A_1, D_{2n+1})$ AD theory, which has exactly $SU(2)$ flavor symmetry \cite{Agarwal:2018zqi}. It would be interesting to find a 5d analog of such AD theories. One natural candidate would be $SU(N)_{N}$ pure YM theory, which has exactly $SU(2)$ flavor symmetry \cite{Bergman:2013aca, Tachikawa:2015mha}. 

\begin{acknowledgments}
The author would like to thank Pedro Liendo for asking a question whether certain short-multiplet vanishes in a mixed OPE of the $H_1$ Argyres-Douglas theory. He also thanks Prarit Agarwal, Ki-Hong Lee, and Hee-Cheol Kim for discussion. He especially thanks Prarit Agarwal for sharing his code constructing various representations of superconformal algebra. 
This work is supported by the National Research Foundation of Korea (NRF) grant NRF-2020R1C1C1007591 and the Start-up Research Grant for new faculty provided by Korea Advanced Institute of Science and Technology (KAIST).

\end{acknowledgments}

\bibliographystyle{jhep}
\bibliography{refs}

\end{document}